\documentclass[%
twocolumn, 
nofootinbib,
 amsmath,amssymb,
prl,
longbibliography
]{revtex4}

\usepackage{graphicx}% Include figure files
\usepackage{dcolumn}% Align table columns on decimal point
\usepackage{bm}% bold math
\usepackage{dsfont}
\usepackage{cases}
\usepackage{color}
    \usepackage{hyperref}
\usepackage{blindtext}
\usepackage{float}
\usepackage{lipsum}
\usepackage{braket}
\usepackage[T1]{fontenc}

\usepackage[]{natbib}

\begin{document}

%\preprint{APS/123-QED}

\title{On polarons and dimerons  in the two-dimensional attractive Hubbard model} 

\author{G. Pascual$^1$\footnote{ \href{mailto:gerard.pascual@upc.edu}{gerard.pascual@upc.edu}}, 
        J. Boronat$^1$, 
        and K. Van Houcke$^2$}

\affiliation{
    $^1$ Departament de F\'{i}sica i Enginyeria Nuclear, Universitat Polit\`ecnica de Catalunya, Campus Nord B4-B5, E-08034, Barcelona, Spain \\
    $^2$ Laboratoire de Physique de l'Ecole Normale Sup\'erieure, ENS, Universit\'e PSL, CNRS, Sorbonne Universit\'e, Universit\'e Paris Cit\'e, F-75005 Paris, France
}

\date{\today}% It is always \today, today,

\begin{abstract}
A two-dimensional spin-up ideal Fermi gas interacting attractively with a spin-down impurity  in the continuum undergoes, at zero temperature, a first-order phase transition from a polaron to a dimeron state. Here we study a similar system on a square lattice, by considering the attractive 2D Fermi-Hubbard model with a single spin-down and a finite filling fraction of spin-up fermions. 
 We study polaron and dimeron quasi-particle properties via variational Ansatz up to one particle-hole excitation. Moreover, we develop a determinant diagrammatic Monte Carlo algorithm for this problem based on expansion in bare on-site coupling $U$. This algorithm turns out to be sign-problem free at any filling of spin-up fermions, allowing one to sample very high diagram order (larger than $200$ in our study) and to do simulations for large $U/t$ (we go up to $U/t=-20$ with $t$ the hopping strength). Both methods give qualitatively consistent results. With variational Ansatz we go to even larger on-site attraction. At very low spin-up filling fraction, we observe the polaron-to-dimeron transition, in agreement with the continuum case. Upon increasing the filling fraction, however, the transition shifts to higher values of $|U|/t$ and the transition disappears beyond a filling fraction of about $20\%$.
In this region, the polaron state always gives a lower energy and has a finite quasi-particle residue.
Our findings are directly relevant to cold atom experiments with 2D optical lattices: a small and finite density of spin-down impurities in the ground state will form a superfluid at strong coupling at low spin-up filling fraction. Above some critical spin-up filling fraction, the system is expected to remain a normal Fermi liquid, even in the strong-coupling limit.

\end{abstract}

\maketitle

\emph{Introduction}--- A  single mobile impurity coupled to a bath of particles is a paradigmatic many-body problem that arises in a large variety of different physical systems, ranging from proton impurities in neutron stars~\cite{Nemeth_1968} to excitons in doped semiconductors~\cite{Sidler_2017}. Typically, the impurity is dressed by excitations of the bath, giving rise to a quasi-particle called polaron~\cite{Landau_1933,Landau_1948}. Such quasi-particles were first proposed by Landau who studied the properties of an electron moving in a crystal lattice, where the coupling of the electron to phonons renormalizes the characteristic properties of the bare electron, such as its energy and mass. 
In the last decades, the creation and unprecedented control of ultracold atomic mixtures has initiated a very active research activity on polaron physics~\cite{Massignan_2014,Schmidt_2018,Scazza_2022,Parish_2023}. 
A key feature is the tunability of the two-body interaction between the impurity and the bath particles via Feshbach resonances. Both baths consisting of fermionic~\cite{Schirotzek_2009,Nascimbene_2009,Kohstall_2012,Koschorreck_2012,Zhang_2012,Wenz_2013,Ong_2015,Cetina_2015,Cetina_2016,Scazza_2017,Mukherjee_2017,Yan_2019,Darkwah_2019,Ness_2020,Fritsche_2021} and bosonic~\cite{Hu_2016,Jorgensen_2016,Yan_2020,Skou_2021} atoms have been realized, corresponding respectively to the so-called Fermi and Bose polarons.

It has been shown by diagrammatic Monte Carlo that the system consisting of a spin-down impurity interacting attractively in the zero-range limit with an ideal spin-up Fermi gas in the continuum undergoes a first-order phase transition in the ground state from polaron (impurity dressed with particle-hole excitations of the sea) to dimeron (bound pair of fermions of opposite spin dressed with particle-hole excitations of the sea) upon increasing the interaction strength, both in three (3D)~\cite{polaron1,polaron2,Vlietinck_3D} and two (2D)~\cite{Vlietinck_2D} spatial dimensions. In one spatial dimension, Bethe Ansatz shows that no sharp transition occurs~\cite{McGuire_1966}.
Remarkably, the ground state of the system can be accurately approximated by a simple Ansatz involving the impurity and a small number of particle-hole excitations~\cite{Chevy_2006, Combescot_2008}. Such Ans\"atze have been used extensively to study the properties of the Fermi polaron in 2D~\cite{Parish_2011,Zollner_2011,Parish_2013,Peng_2021} and 3D~\cite{Punk_2009,Mora_2009,Bruun_2010,Schmidt_2011,Mathy_2011,Trefzger_2012,Cui_2020,Peng_2021}. 

On the lattice, a mobile impurity moving within a 2D Bose-Hubbard bath at zero temperature was recently studied, revealing a well-defined polaron throughout almost the entire phase diagram of the bath~\cite{Colussi_2023,Santiago_2024}. 
For fermions, the Fermi polaron problem was considered within the Fermi-Hubbard model 
in an early study by Sorella~\cite{Sorella_1994}, where the quasi-particle residue was computed (for $U/t = -4, -\infty$) using quantum Monte Carlo, second-order perturbation theory and/or a variational wave function introduced in \cite{Edwards_1990}. Sorella concluded that the polaron residue ``looks always finite", except for a half filled band. 
Studying impurities in the Hubbard model with different configurations for the background is an active field of research~\cite{Hu_2024,Amelio_2024a,Amelio_2024b}.
For $U>0$, magnetic polarons~\cite{Grusdt_2018,Grusdt_2019,Blomquist_2020,Nielsen_2021} have been observed in the doped Hubbard model~\cite{Koepsell_2019}. 
Experimentally, ultracold atoms in optical lattices provide a platform to realize Hubbard models~\cite{Jaksch_1998,Gross_2017}. 
In particular, quantum gas microscopes allow the observation and control of quantum matter at the level of individual atoms~\cite{Bakr_2009,Sherson_2010,Cheuk_2015,Haller_2015,Parsons_2015, Omran_2015,Edge_2015,Verstraten_2024}, thus providing an ideal observation tool of lattice systems.

In this Letter, we focus on the Fermi-polaron problem on the lattice, by considering the 2D attractive Hubbard model with strong spin-population imbalance, i.e., a single spin-$\downarrow$ and a finite number $N_\uparrow$ of spin-$\uparrow$ fermions. We calculate the polaron and dimeron properties by variational Ansatz and Diagrammatic Monte Carlo.
Our work extends the study of the Fermi polaron to include an (optical) lattice, revealing a fundamental difference from the well-studied continuum case: the polaron-dimeron transition disappears above a critical filling fraction. This qualitatively new behavior could be observed in experiments on ultra-cold fermionic atoms in a 2D optical lattice.

\emph{Model}--- We consider the attractive Hubbard model for a 2D square lattice with $N$ sites ($N=L\times L$, where $L$ is the number of sites per dimension) filled with $N_{\uparrow}$ spin-up fermions and a single spin-down impurity. The Hamiltonian reads
\begin{eqnarray}\label{eq: Hamiltonian}
\hat{H}  =   -t \sum_{\sigma}\sum_{\langle \mathbf{r}, \mathbf{r}'\rangle} \left( \hat{c}^{\dagger}_{\mathbf{r},\sigma}
\hat{c}^{\phantom{\dagger}}_{\mathbf{r}',\sigma} + h.c. \right) + U \sum_{\mathbf{r}} \hat{n}_{\mathbf{r},\uparrow} \hat{n}_{\mathbf{r},\downarrow}  \; ,
\end{eqnarray}
where $\sigma=\uparrow,\downarrow$, $\hat{c}^{\dagger}_{\mathbf{r},\sigma}$ creates a spin-$\sigma$ fermion on site $\mathbf{r}$, and $\hat{n}_{\mathbf{r},\sigma} = \hat{c}^{\dagger}_{\mathbf{r},\sigma}\hat{c}^{\phantom{\dagger}}_{\mathbf{r},\sigma}$ the on-site number operator. 
The first term describes hopping of fermions between nearest-neighbor sites with amplitude $t$,  while the second term describes on-site attraction with amplitude $U<0$. 
Note that the repulsive model is mapped onto the attractive one after particle-hole transformation of one spin-component. 
We assume periodic boundary conditions and we set the lattice spacing to $1$.

\emph{Methods}--- To determine the properties of the lattice Fermi polaron, we first consider a variational Ansatz similar to the one for the continuum problem~\cite{Chevy_2006,Cui_2020,Peng_2021},
\begin{equation}\label{Ansatz_Polaron}
\begin{split}
\ket{P(\mathbf{Q}_P)} = \: &\alpha_0\hat{c}^{\dag}_{\mathbf{Q}_P,\downarrow}\ket{\text{FS}_{N_{\uparrow}}} +\\ &\sum_{\mathbf{k},\mathbf{q}}{}^{'}\alpha_{\mathbf{k},\mathbf{q}}\:\hat{c}^{\dag}_{\mathbf{Q}_P+\mathbf{q}-\mathbf{k},\downarrow}\:\hat{c}^{\dag}_{\mathbf{k},\uparrow}\hat{c}_{\mathbf{q},\uparrow}\ket{\text{FS}_{N_{\uparrow}}} \; ,
\end{split}
\end{equation}
where $\hat{c}^{\dag}_{\mathbf{k},\sigma}$ ($\hat{c}_{\mathbf{k},\sigma}$) creates (annihilates) a fermion with quasi-momentum $\mathbf{k}$, 
and $\ket{\text{FS}_{N_{\uparrow}}}$ is the Fermi sea of $N_{\uparrow}$ spin-up particles, i.e., the ground state of the non-interacting bath. The particle number 
$N_{\uparrow}$ is always chosen such that $\ket{\text{FS}_{N_{\uparrow}}}$ is a non-degenerate, translationally invariant closed-shell state.
In that case, $\mathbf{Q}_P$ is the total quasi-momentum of the polaron state. %, since $\ket{\text{FS}_{N_{\uparrow}}}$ has zero quasi-momentum. 
The prime in Eq.~(\ref{Ansatz_Polaron}) indicates that sums on $\mathbf{k}$ and $\mathbf{q}$ are restricted to quasi-momenta above and below the Fermi level, respectively.
In addition to the Ansatz~(\ref{Ansatz_Polaron}) we also consider an Ansatz for the dimeron state~\cite{Punk_2009,Combescot_2009,Mora_2009,Parish_2011,Parish_2013,Cui_2020,Peng_2021}: 
\begin{equation}\label{Ansatz_Molecule}
\begin{split}
& \ket{M(\mathbf{Q}_M)} = \sum_{\mathbf{k}}{}^{'}\xi_{\mathbf{k}}\hat{c}^{\dag}_{\mathbf{Q}_M-\mathbf{k},\downarrow}\hat{c}^{\dag}_{\mathbf{k},\uparrow}\ket{\text{FS}_{N_{\uparrow}-1}^{(\mathbf{Q}_{\text{FS}})}}  \\  & + \frac{1}{2}\sum_{\mathbf{k},\mathbf{k'},\mathbf{q}}{}^{'}\xi_{\mathbf{k},\mathbf{k}',\mathbf{q}}\:\hat{c}^{\dag}_{\mathbf{Q}_M+\mathbf{q}-\mathbf{k}-\mathbf{k}',\downarrow}\:\hat{c}^{\dag}_{\mathbf{k},\uparrow}\hat{c}^{\dag}_{\mathbf{k}',\uparrow}\hat{c}_{\mathbf{q},\uparrow}\ket{\text{FS}_{N_{\uparrow}-1}^{(\mathbf{Q}_{\text{FS}})}} \; ,
\end{split}
\end{equation}
where the Fermi sea %in Eq.~(\ref{Ansatz_Molecule}) 
has $N_{\uparrow}-1$ particles such that both Ans\"atze (\ref{Ansatz_Polaron}) and (\ref{Ansatz_Molecule}) consider the same total number of fermions. 
Since all the levels of the non-interacting spectrum have a degeneracy of $4$ or $8$ (apart from $\mathbf{k} = \mathbf{0}$), the Fermi sea with $N_{\uparrow}-1$ particles
is degenerate and, as a result, there are multiple equivalent Ans\"atze.  
We fix the total quasi-momentum of the Fermi sea to $\mathbf{Q}_{\text{FS}}$, and write $| \text{FS}_{N_{\uparrow}-1}^{(\mathbf{Q}_{\text{FS}})}\rangle = \hat{c}^{\phantom{\dagger}}_{-\mathbf{Q}_{\text{FS}},\uparrow}|\text{FS}_{N_{\uparrow}}\rangle$. 
The total quasi-momentum of the dimeron state is  $\mathbf{Q}_{\text{FS}}+\mathbf{Q}_M$.
Again, the prime implies summing $\mathbf{k}$ and $\mathbf{k}'$ ($\mathbf{q}$) above (below) the Fermi level. 
A priori, we wish to compute the energy of quasi-particles with zero center-of-mass momentum, i.e. $\mathbf{Q}_P=0$ for the polaron and $\mathbf{Q}_M=0$ for the dimeron. 
However, keeping the possibility of setting $\mathbf{Q}_P$ and $\mathbf{Q}_M$ to the Fermi wave vector will enable a unified variational Ansatz~\cite{Cui_2020,Peng_2021}. 

By minimizing the energy 
for both Ans\"atze, as was done in the continuum in \cite{Chevy_2006,Punk_2009,Parish_2011,Peng_2021}, one obtains the following equations for the polaron,

\begin{equation}\label{eq_1_Polaron}
     \left(E-E_{\mathbf{Q}_P}^{(0)}\right)\alpha_0 = \frac{U}{N}\sum\limits_{\mathbf{k},\mathbf{q}}{}^{'}\alpha_{\mathbf{k},\mathbf{q}} \:,
\end{equation}
\begin{equation}\label{eq_2_Polaron}
    \left(E-E_{\mathbf{Q}_P,\mathbf{k},\mathbf{q}}^{(1)}\right)\alpha_{\mathbf{k},\mathbf{q}} = \frac{U}{N}\alpha_0 + \frac{U}{N}\sum\limits_{\mathbf{k}'}{}^{'}\alpha_{\mathbf{k}',\mathbf{q}} - \frac{U}{N}\sum\limits_{\mathbf{q'}}{}^{'}\alpha_{\mathbf{k},\mathbf{q'}} \:,
\end{equation}
where $E_{\mathbf{Q}_P}^{(0)} = E_{\text{FS}}^{N_{\uparrow}} + \epsilon_{\mathbf{Q}_P} + U\frac{N_{\uparrow}}{N}$ and $E_{\mathbf{Q}_P,\mathbf{k},\mathbf{q}}^{(1)} = E_{\text{FS}}^{N_{\uparrow}} + \epsilon_{\mathbf{k}} - \epsilon_{\mathbf{q}} + \epsilon_{\mathbf{Q}_P+\mathbf{q}-\mathbf{k}} + U\frac{N_{\uparrow}}{N}$, with $E_{\text{FS}}^{N_{\uparrow}}$ the energy of the Fermi sea with $N_{\uparrow}$ spin-up fermions and $\epsilon_{\mathbf{k}}=-2t(\cos(k_x)+\cos(k_y))$; and, for the molecule,

\begin{equation}\label{eq_1_Molecule}
\begin{split}
    \left(E-E_{\mathbf{Q}_M,\mathbf{k}}^{(2)}\right)\xi_{\mathbf{k}} = \frac{U}{N}\sum\limits_{\mathbf{k}'}{}^{'}\xi_{\mathbf{k}'} + \frac{U}{N}\sum\limits_{\mathbf{k}''\mathbf{q}}{}^{'}\xi_{\mathbf{k},\mathbf{k}'',\mathbf{q}} \:,
\end{split}
\end{equation}
\begin{equation}\label{eq_2_Molecule}
\begin{split}
    \left(E-E_{\mathbf{Q}_M,\mathbf{k},\mathbf{k}',\mathbf{q}}^{(3)}\right)\xi_{\mathbf{k},\mathbf{k}',\mathbf{q}} = \frac{U}{N}\left(\xi_{\mathbf{k}} - \xi_{\mathbf{k}'}\right)\\+\frac{U}{N}\sum\limits_{\mathbf{k}''}{}^{'}\xi_{\mathbf{k},\mathbf{k}'',\mathbf{q}}  + \frac{U}{N}\sum\limits_{\mathbf{k}''}{}^{'}\xi_{\mathbf{k}'',\mathbf{k}',\mathbf{q}} - \frac{U}{N}\sum\limits_{\mathbf{q''}}{}^{'}\xi_{\mathbf{k},\mathbf{k}',\mathbf{q''}} \: ,
\end{split}
\end{equation}
where $E_{\mathbf{Q}_M,\mathbf{k}}^{(2)} = E_{\text{FS}}^{N_{\uparrow}-1} + \epsilon_{\mathbf{k}} + \epsilon_{\mathbf{Q}_M-\mathbf{k}} + U\frac{N_{\uparrow}-1}{N}$ and $E_{\mathbf{Q}_M,\mathbf{k},\mathbf{k}',\mathbf{q}}^{(3)} = E_{\text{FS}}^{N_{\uparrow}-1} + \epsilon_{\mathbf{k}} + \epsilon_{\mathbf{k}'} - \epsilon_{\mathbf{q}} + \epsilon_{\mathbf{Q}_M+\mathbf{q}-\mathbf{k}-\mathbf{k}'} + U\frac{N_{\uparrow}-1}{N}$. 
Note that, in contrast to the continuum case, there is a finite bandwidth. Moreover, some 
of the terms in Eqs.~(\ref{eq_1_Polaron}), (\ref{eq_2_Polaron}), (\ref{eq_1_Molecule}) and (\ref{eq_2_Molecule}) vanish  in the continuum limit. % in the thermodynamic limit. 

Terms in Eqs. (\ref{eq_1_Polaron}) and (\ref{eq_2_Polaron}) can be rearranged to transform the system of equations into a single homogeneous Fredholm equation of the second kind~\cite{Punk_2009}. The ground-state energy is computed by finding the value of $E$ for which the determinant of the kernel vanishes. The same method is applied to the system of Eqs. (\ref{eq_1_Molecule}) and (\ref{eq_2_Molecule}) but at a higher computational cost.
Alternatively, the above equations can be solved using the iterative method. We have checked that this produces consistent results. 

In addition to the variational calculations, we also performed diagrammatic Monte Carlo simulations for the lattice polaron problem, in order to determine
the effect of higher number of particle-hole excitations. 
To study the ground-state properties, we consider the single-particle propagator:
\begin{eqnarray}
G_{\downarrow}(\mathbf{k},\tau) =  - \theta(\tau) \bra{\text{FS}_{N_{\uparrow}}} 
\hat{c}^{\phantom{\dagger}}_{\mathbf{k},\downarrow}(\tau) \hat{c}^{\dagger}_{\mathbf{k},\downarrow}(0)\ket{  \text{FS}_{N_{\uparrow}}},
\label{eq:Gdef}
\end{eqnarray} 
where operators are written in the imaginary-time Heisenberg picture, $\hat{c}^{\phantom{\dagger}}_{\mathbf{k},\downarrow}(\tau) = e^{\hat{K}\tau} \hat{c}_{\mathbf{k},\downarrow}  e^{-\hat{K}\tau}$ with $\hat{K}=\hat{H} -  \sum\limits_\sigma \mu_{\sigma} \hat{N}_{\sigma}$, and with $\mu_{\sigma}$ the chemical potentials for both spin components. 
We developed and implemented  a polaron determinant (PDet) algorithm~\cite{pdet} for evaluating the diagrammatic expansion in terms of bare $U$ for the propagator $G_{\downarrow}$.

The algorithm works in position-imaginary time representation, where the sum of all diagram topologies for a given set of space-time coordinates of the interaction vertices is given by a single determinant~\cite{Rubtsov_2003,Rubtsov_2004,Burovski_2004}. For the polaron problem,
all diagrams generated in this way are automatically connected~\cite{pdet}.  The expansion of the polaron propagator in powers of $U$ is written as  
\begin{eqnarray}
    G_{\downarrow}(X) & =  G_{\downarrow}^{0}(X) + \sum\limits_{n=1}^{\infty}
    \int_{X_1}\ldots \int_{X_n} 
   U^n G_{\downarrow}^{0}(X_1)   \nonumber \\
  &   G_{\downarrow}^{0}(X_2-X_1) \cdots G_{\downarrow}^{0}(X-X_n)  ~\det(\mathcal{M}^{(n)}) \; ,
\label{eq:Gexpansion}
\end{eqnarray}
with $\mathcal{M}^{(n)}$ an $n\times n$ matrix with elements $\mathcal{M}^{(n)}_{i,j} = G^{0}_{\uparrow}(X_i-X_j)$, with $i,j=1,\ldots,n$, where $n$ is the order of the diagrams
We have used the notation $X = (\mathbf{r},\tau)$ and $\int_X := \sum\limits_{\mathbf{r}}\int_0^{+\infty} d\tau$. 
The $G^{0}_{\sigma}$ are the non-interacting spin-$\sigma$ single-particle propagators.  The different terms in Eq.~(\ref{eq:Gexpansion}) are evaluated 
stochastically by the Metropolis algorithm~\cite{Pascual_2025}.

Extracting the quasi-particle energy from $G_{\downarrow}$ calculated by Monte Carlo typically requires fitting the propagator at large enough~$\tau$~\cite{polaron1,polaron2,Vlietinck_3D}. We propose here a new direct method to estimate the energy, without fitting and without calculating the self-energy. We follow a standard approach for quantum Monte Carlo calculations: given a trial state $|\psi_T\rangle$ for the system,
the operator $e^{-\hat{K}\tau}$ acts as a projector onto the ground state:
\begin{eqnarray}
\frac{\bra{\psi_T}  e^{-\hat{K}\tau/2} \hat{H} e^{-\hat{K}\tau/2} \ket{\psi_T}}{\bra{\psi_T}   e^{-\hat{K}\tau} \ket{\psi_T}} \overset{\tau\to+\infty}{=} E \; .
\label{eq:projector}
\end{eqnarray} 
This will yield the true ground-state energy as long as $\braket{\psi_T | \Psi_0^{N_\uparrow}} \neq 0$, with $|  \Psi_0^{N_\uparrow}  \rangle$ the ground state of the interacting system.
If we choose the trial state  to be $ | \psi_T\rangle =  \hat{c}^{\dagger}_{\mathbf{k},\downarrow}| \text{FS}_{N_{\uparrow}} \rangle$, which is an eigenstate of the non-interacting Hamiltonian, then the energy estimator in the l.h.s. of Eq.~(\ref{eq:projector}) gives the energy
\begin{eqnarray}
E^{N_\uparrow}(\mathbf{k},\tau) &  \equiv & E^{N_\uparrow}_{\text{FS}} + \epsilon_{\mathbf{k},\downarrow}   \nonumber \\
&  +  &\frac{\bra{\text{FS}_{N_{\uparrow}}} \hat{c}^{\phantom{\dagger}}_{\mathbf{k},\downarrow}\hat{V}e^{-\hat{K}\tau}\hat{c}^{\dagger}_{\mathbf{k},\downarrow}\ket{\text{FS}_{N_{\uparrow}}}}
{\bra{\text{FS}_{N_{\uparrow}}} \hat{c}^{\phantom{\dagger}}_{\mathbf{k},\downarrow} e^{-\hat{K}\tau} \hat{c}^{\dagger}_{\mathbf{k},\downarrow} \ket{\text{FS}_{N_{\uparrow}}}} \; ,
\label{eq:estimatorE}
\end{eqnarray}
with $\hat{V}$ the on-site Hubbard interaction term and $\mathbf{k}$ the total quasi-momentum. 
The denominator is proportional to $G_{\downarrow}(\mathbf{k},\tau)$, which is calculated with PDet based on Eq.~(\ref{eq:Gexpansion}). 
To evaluate the numerator, one needs to consider diagrams having an extra interaction  vertex at imaginary time $\tau^-$. This can be achieved in the Monte Carlo algorithm by extending the configuration space with an extra sector where diagrams have such an extra interaction vertex. The numerator and denominator in Eq.~(\ref{eq:estimatorE}) are both calculated within the same simulation, by enabling the Monte Carlo algorithm to switch between the different sectors that evaluate them.

\begin{figure}[h!]
\centering
\includegraphics[scale=0.95]{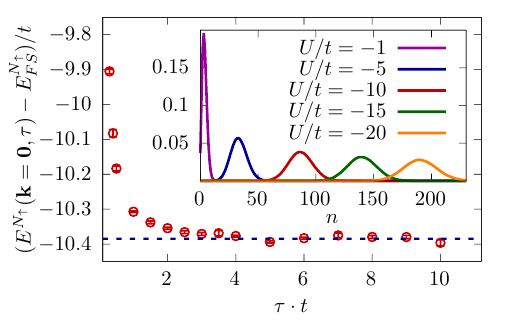}
\caption{Monte Carlo estimate for the polaron quasi-particle energy $E_P$ as function of the imaginary time $\tau$ for $U/t=-10$, showing convergence to the ground state at large enough $\tau$. The dashed line represents a fit to the plateau.
The inset shows histograms of the sampled diagram order $n$ for different values of $U/t$ at $\tau t=10$. Simulations were performed for $L=23$ and $N_{\uparrow} = 157$.
}
\label{fig:tau_E}
\end{figure}

\emph{Results}--- We have performed PDet simulations for the lattice Fermi polaron up to coupling $U/t=-20$. Fig.~\ref{fig:tau_E} shows the estimator 
$E^{N_\uparrow}(\mathbf{k}=\mathbf{0},\tau) - E^{N_\uparrow}_{\text{FS}}$ (see Eq.~(\ref{eq:estimatorE})) for the polaron energy $E_P$
for $U/t=-10$ as a function of imaginary time $\tau$.  At large enough $\tau$ convergence to the  ground-state energy is observed.
When performing PDet simulations for the lattice polaron, we find that  \emph{all} configurations sampled during the Monte Carlo process have the \emph{same sign}. This remarkable fact allows us to go to very high order, and thus large values of $U/t$. The inset of Fig.~\ref{fig:tau_E} shows histograms of the order $n$ for different values of $U/t$ when calculating the denominator of Eq.~(\ref{eq:estimatorE}) at $\tau t=10$ and $\mathbf{k} = 0$. For $U/t=-20$ we go to diagram orders beyond 200. For the attractive Hubbard model, having $\rho_{\downarrow} = \rho_{\uparrow}$ prevents the sign problem from occurring~\cite{Burovski_2004}. In our case of a single spin-down and an arbitrary filling of the spin-up component, however, there is no such obvious symmetry. 

The dimeron properties can be extracted from the large-time behavior of the two-particle Green's function 
$G_{\uparrow\downarrow}(\mathbf{r},\tau)  =  - \theta(\tau) \bra{\text{FS}_{N_{\uparrow}-1}} 
(\hat{c}^{\phantom{\dagger}}_{\mathbf{r},\downarrow}\hat{c}^{\phantom{\dagger}}_{\mathbf{r},\uparrow})(\tau) (\hat{c}^{\dagger}_{\mathbf{0},\uparrow}\hat{c}^{\dagger}_{\mathbf{0},\downarrow})(0)\ket{  \text{FS}_{N_{\uparrow}-1}} $~\cite{polaron1,polaron2,Vlietinck_3D}.  
$G_{\uparrow\downarrow}$ can be calculated via a PDet algorithm similar to $G_{\downarrow}$. However,  we find that the diagrammatic contributions to $G_{\uparrow\downarrow}$ do change their sign, making it challenging to go to the high diagram orders needed at large values of $\tau$. We therefore use a different approach to get the dimeron energy.  As shown in Refs.~\cite{Cui_2020,Peng_2021}, the polaron-to-dimeron transition in the continuum case can be viewed as a crossing of polaron Ansatz states with different momenta, $\mathbf{Q}=0$ and $|\mathbf{Q}| = k_F$, with $\mathbf{Q}$ defined as the momentum of the Fermi polaron with respect to the Fermi sea and with $k_F$ the Fermi momentum.
Analogously, we determine the energy of the dimeron state by considering the Fermi polaron propagator with a finite quasi-momentum at the Fermi level. More precisely, we use Eq.~(\ref{eq:estimatorE}) with $\mathbf{k}$ at the Fermi surface. 

\begin{figure}
\centering
\includegraphics[scale=0.95]{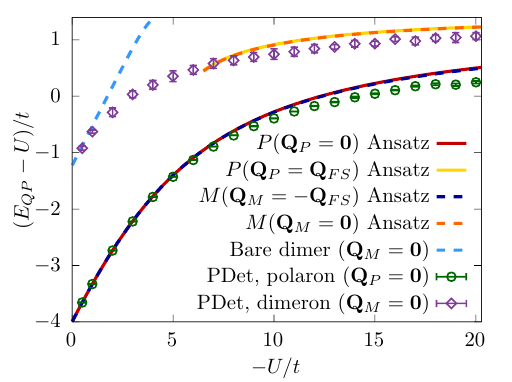}
\caption{The quasiparticle energy $E_{QP}$ (shifted by $U$) as function of the coupling. PDet Monte Carlo results for the polaron and dimeron are shown with circles and diamonds, respectively.  Energies
based on variational Ans\"atze (\ref{Ansatz_Polaron}) and (\ref{Ansatz_Molecule}) are shown with lines. Calculations were done for $L=23$ and 
$N_{\uparrow} = 157$.}
%$\rho_{\uparrow} \simeq 0.29$.}
\label{fig:E}
\end{figure}

In Fig.~\ref{fig:E} we show the quasi-particle energies $E_{QP} = E - E^{N_{\uparrow}}_{\text{FS}}$ (shifted by $U$ for convenience) of polaron and dimeron branches obtained by variational Ansatz and diagrammatic Monte Carlo for filling $\rho_{\uparrow} \equiv N_{\uparrow}/N \simeq 0.29$ and $L = 23$. We have checked that finite-size effects are negligible at this lattice size and filling.
The green circles and purple diamonds  represent Monte Carlo estimates for the polaron  energy
 $E_{QP}\equiv E_P = E^{N_{\uparrow}}(\mathbf{k} = \mathbf{0},\tau=\infty) - E^{N_{\uparrow}}_{\text{FS}}$
 and the dimeron energy
 $E_{QP}\equiv E_D = E^{N_{\uparrow}}(\mathbf{k} = \mathbf{Q}_{\text{FS}},\tau=\infty) - E^{N_{\uparrow}}_{\text{FS}}$, respectively. 
The red solid line shows the polaron energy from the variational Ansatz $\ket{P(\mathbf{Q}_P=\mathbf{0})}$ (see Eq.~(\ref{Ansatz_Polaron})). Alternatively, one can also get the polaron energy  via the Ansatz $\ket{M(\mathbf{Q}_M=-\mathbf{Q}_{\text{FS}})}$~\footnote{Indeed, one notes that the specific choice
$\xi_{\mathbf{k}} = \alpha_0 \delta_{\mathbf{k},-\mathbf{Q}_{\text{FS}}} + \alpha_{\mathbf{k},-\mathbf{Q}_{\text{FS}}} \;$ and
$\xi_{\mathbf{k},\mathbf{k}',\mathbf{q}} = 2 \alpha_{\mathbf{k}',\mathbf{q}} ~ \delta_{\mathbf{k},-\mathbf{Q}_{\text{FS}}}$ 
reduces the Ansatz (\ref{Ansatz_Molecule}) with $\mathbf{Q}_M=-\mathbf{Q}_{\text{FS}}$ to $\ket{P(\mathbf{Q}_P = \mathbf{0})}$.}.
This gives a slighly lower energy at high values of $|U|/t$ than Ansatz $\ket{P(\mathbf{Q}_P=\mathbf{0})}$.
The orange dashed line shows the dimeron energy from the variational Ansatz $\ket{M(\mathbf{Q}_M=\mathbf{0})}$ (see Eq.~(\ref{Ansatz_Molecule})). For $|U|/t \lesssim 6$, we do not find any solutions. The yellow solid line corresponds to the dimeron energy obtained with the Ansatz $\ket{P(\mathbf{Q}_P=\mathbf{Q}_{\text{FS}})}$~\footnote{Here, note that the specific choice
$\alpha_0 = \xi_{-\mathbf{Q}_{\text{FS}}}$ and
$\alpha_{\mathbf{k},\mathbf{q}} = \xi_{\mathbf{k}} \delta_{\mathbf{q},-\mathbf{Q}_{\text{FS}}}$
 in (\ref{Ansatz_Polaron}) reduces $\ket{P(\mathbf{Q}_P=\mathbf{Q}_{\text{FS}})}$ to the first term of $\ket{M(\mathbf{Q}_M=\mathbf{0})}$.}.
The $E_D$ from PDdet joins the variational energy calculated with first term of $\ket{M(\mathbf{Q}_M=\mathbf{0})}$, to which we refer as the \emph{bare dimer} solution, at small enough $U/t$. 
 At $U=0$ one has  $E_D=\epsilon(\mathbf{Q}_{\text{FS}}) \simeq -1.2t$ for the filling fraction considered in Fig.~\ref{fig:E}. 
 We conclude from Fig.~\ref{fig:E} that there is no sharp polaron-to-dimeron transition for $|U|/t \leq 20$. Using the different variational Ans\"atze, we went beyond $U/t=-20$
 and did not find any sharp transition from polaron to dimeron at any coupling.

\begin{figure}[!]
	\centering
	\includegraphics[]{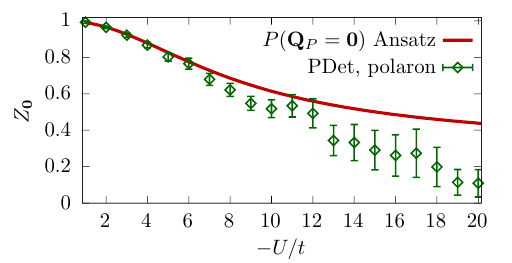}
	\caption{Quasi-particle residue $Z_{\mathbf{0}}$ as function of $U$ at $L=23$ and $\rho_{\uparrow} \simeq 0.29$, calculated with variational Ansatz $\ket{P(\mathbf{Q}_P=\mathbf{0})}$ and PDet.}
    \label{fig:residue}
\end{figure}

In Fig. \ref{fig:residue} we show the quasi-particle residue $Z_{\mathbf{0}} \equiv | \bra{ \Psi_0^{N_{\uparrow}}}  \hat{c}^{\dagger}_{\mathbf{0},\downarrow} \ket{\text{FS}_{N_{\uparrow}}}|^2$ for filling $\rho_{\uparrow} \simeq 0.29$. When the coupling increases the residue drops, indicating the greater importance of particle-hole excitations. Monte Carlo and polaron Ansatz results agree quantitatively well for $|U|/t \lesssim 6$. We find that $Z$ remains finite at all couplings for the polaron Ansatz.  
The polaron residue was obtained in the Monte Carlo algorithm via $G_{\downarrow}(\mathbf{k}=\mathbf{0},\tau) \overset{\tau\to+\infty}{=} - Z_{\mathbf{0}} e^{-(E_P-\mu_\downarrow)\tau}$. 

\begin{figure}[!]
	\centering
	\includegraphics[]{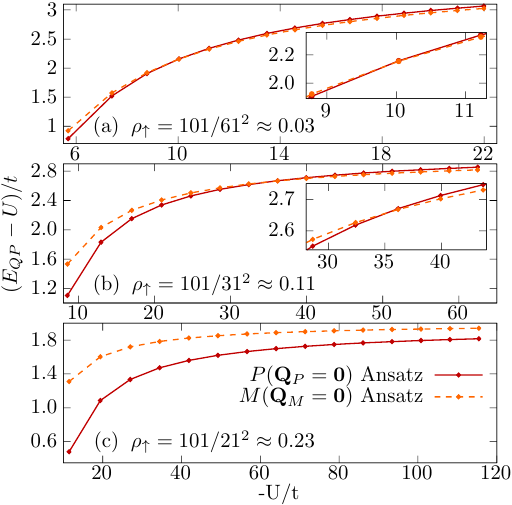}
	\caption{Quasi-particle energies as function of $-U/t$ for different filling fractions $\rho_{\uparrow} = N_{\uparrow}/L^2$, obtained by the variational Ansatz approach.
	The inset magnifies the region of $-U/t$ where the polaron-dimeron transition occurs. }
    \label{fig:fillings}
\end{figure}

Finally, in Fig. \ref{fig:fillings} we show the polaron and dimeron quasi-particle energies for different filling factors $\rho_{\uparrow}$ using the variational approach. 
At low enough  $\rho_{\uparrow}$, we observe the polaron-dimeron transition, consistent with the continuum case~\cite{Vlietinck_2D}. 
Upon increasing  $\rho_{\uparrow}$ the critical $-U/t$ shift to higher values and, ultimately, the transition disappears above a critical $\rho_{\uparrow} \simeq 0.2$. 

One might ask why the transition disappears in the lattice beyond some filling fraction. An upper bound for the critical filling fraction can be obtained as follows. The dimeron energy is bounded by $E_D - U  \geq \epsilon_F$ with $\epsilon_F \equiv \epsilon(\mathbf{Q}_{\text{FS}})$ the Fermi energy.
In addition, an upper bound to the polaron energy is obtained via the polaron Ansatz with following homogeneous choice for the variational parameters:
$\alpha_0=  \sqrt{\rho_\uparrow}$ and  $\alpha_{\mathbf{k},\mathbf{q}} = 1/(N\sqrt{\rho_{\uparrow}})$. This gives the upper bound  $E_P - U \leq -4t  \rho_{\uparrow} +  (1-\rho_{\uparrow})12t$ for the exact polaron energy $E_P$. The two bounds are equal for $\rho_{\uparrow} = \frac{3}{4}  - \frac{\epsilon_F}{16t} \simeq 0.7$, thus showing that there can be no polaron-dimeron transition beyond this filling fraction. The bounds are essentially caused by Pauli-blocking and the reduced available phase space for particle-hole excitations upon increasing $\rho_\uparrow$. 
Moreover, a particle-hole transformation for the spin-up component with density $\rho_\uparrow = 1 - \eta$ maps the attractive model onto the repulsive one with a small density of $\eta$ for spin-up. For the repulsive model, there is no dimeron ground state in this dilute regime.

\emph{Conclusion and outlook}--- We have studied the quasi-particle properties of the Fermi polaron in a 2D lattice. To this end, we have developed a \emph{sign-problem-free} PDet Monte Carlo algorithm based on the diagrammatic expansion in bare $U$, with a direct estimator for the quasi-particle energy. 
In addition, we have done calculations based on variational Ans\"atze.
At low enough filling, we observe the polaron-dimeron transition, similar to the continuum case. However, this transition disappears upon increasing the filling fraction beyond a critical value close to $0.2$. 
The absence of the transition is closely related to the increased Pauli-blocking and the reduced available phase space for particle-hole excitations when increasing $\rho_{\uparrow}$. 
Our results are relevant for the ground-state phase diagram of the strongly polarized Fermi gas on the lattice. A small and finite density of spin-down impurities in the ground state will form a superfluid at strong coupling when $\rho_\uparrow$ is low enough. Above some spin-up filling fraction, however, the system will remain a normal Fermi liquid, even in the strong-coupling limit. Determining the precise phase diagram of the highly spin-polarized system would be an interesting topic of future research. 
Our findings are thus directly relevant to cold atom experiments with 2D optical lattices~\cite{Hartke_2023}, where impressive progress has been made recently to reach low temperatures~\cite{Greiner_2025,Chalopin_2025,Wang_2025}. In particular, the stability of the Fermi polaron at high values of $|U|/t$ at filling fractions $\rho_{\uparrow} \gtrsim 0.2$ could be verified experimentally. 
We have also calculated the polaron residue, which remains finite, in agreement with early calculations by Sorella~\cite{Sorella_1994}.
The fact that our PDet algorithm is free of any sign problem was unexpected since there is no obvious symmetry guaranteeing a fixed sign. The implications for simulating a finite spin-down density is a topic we leave for future research.

\begin{acknowledgments}
 \noindent \textit{Acknowledgments.}
 We gratefully acknowledge fruitful discussion with F\'elix Werner and Xavier Leyronas. 
This work has been supported by the French Agence Nationale de la Recherche (ANR) under grant ANR-21-CE30-0033 (project LODIS), by the Spanish Ministerio de Universidades under the grant FPU No. FPU20/00013, by the Spanish Ministerio de Ciencia e Innovaci\'on (MCIN/AEI/10.13039/501100011033, grant PID2023-147469NB-C21), and by the Generalitat de Catalunya (grant 2021 SGR 01411).
\end{acknowledgments}

%\nocite{*}

%\appendix*
%\section{Appendix: Limit to the continuum}
\bibliographystyle{unsrtnat}
\bibliography{main}% Produces the bibliography via BibTeX.

\end{document}